\documentstyle[12pt,epsfig]{article}
\input epsf
\begin{document}

\author{L. Martina, Kur. Myrzakul$^{a)}$, R. Myrzakulov$\thanks{%
Permanent address: Institute of Physics and Technology, 480082, Alma-Ata-82,
Kazakhstan. E-mail: cnlpmyra@satsun.sci.kz}$ and G. Soliani \and %
Dipartimento di Fisica dell'Universit\`{a} and \and INFN-Sezione di Lecce,
73100 Lecce, Italy \and $^{a)}$Institute of Mathematics, Alma-Ata, Kazakhstan}
\title{{\LARGE Deformation of surfaces, integrable systems and 
Chern-Simons theory}}

\maketitle

\begin{abstract}
A few years ago, some of us devised a method to obtain integrable systems in
(2+1)-dimensions from the classical non-Abelian pure Chern-Simons action via
reduction of the gauge connection in Hermitian symmetric spaces. In this
paper we show that the methods developed in studying classical non-Abelian
pure Chern-Simons actions, can be naturally implemented by means of a
geometrical interpretation of such systems. The Chern-Simons equation of
motion turns out to be related to time evolving 2-dimensional surfaces in
such a way that these deformations are both locally compatible with the
Gauss-Mainardi-Codazzi equations and completely integrable. The properties
of these relationships are investigated together with the most relevant
consequences. Explicit examples of integrable surface deformations are
displayed and discussed.
\end{abstract}

\section{Introduction}

Many authors have extensively studied the deep relations among completely
integrable systems and the basic equations of the differential geometry,
like the Frenet formulae defining curves embedded in ${\bf R}^{3}$, or their
analogous for the surfaces, the Gauss-Weingarten (GW) equations, and the
corresponding integrability conditions, i.e. the Gauss-Mainardi-Codazzi
(GMC) equations (see for instance [1,2,3]). In these approaches the main
idea is to add to a generic differential geometry setting certain auxiliary
conditions, containing from the beginning the properties of the completely
integrable systems.

A slightly different situation occurs in the study of the so-called Darboux
system [4,5], which naturally arises in looking for classes of orthogonal
curvilinear coordinates in Euclidean spaces and whose integrability has been
detected in Ref. 6. Such a system has been investigated mainly in connection
with the Topological Field Theory [7]. On the other hand, some years ago
some of us proposed a simple method to obtain completely integrable systems
in ($2+1)$-dimensions, from classes of non Abelian Chern-Simons (CS) field
theories, taking values in Hermitian symmetric spaces [8]. In this context
completely integrable systems are seen as particular gauge choices in which
the theory is formulated. Moreover, linear spectral problems are naturally
related to the geometrical constraints imposed on the target space. From
this point of view, integrable systems arise as special reductions, which
break the general covariance and the gauge invariance of the original field
theory, but preserve a residual symmetry in order to allow the Lax
representation and the complete integrability, although the solvability is
lost.

\bigskip

In the present work we show that this approach can be naturally implemented
by resorting to a geometrical interpretation of the completely integrable
systems mentioned above. Precisely, we show as the CS equation of motion can
describe time evolving 2-dimensional surfaces in such a way that the
deformation is not only locally compatible with the GMC equation, but
completely integrable as well. The nature and the properties of such
relationships are investigated together with the most important
consequences. Furthermore, explicit examples of integrable deformations of
surfaces are displayed.

The paper is organized as follows. Section 2 contains some results on the CS
theory. In Section 3 the fundamental terminology and notations related to
the theory of 2-dimensional surfaces are reviewed. In Section 4 the general
formulation of deformation of 2-dimensional surfaces is presented. Section 5
is devoted to the analysis of certain spin models in (2+1)-dimensions.
Section 6 is addressed to the bilinear representations of the spin systems
fields and of the trihedral moving frame. Sections 7 and 8 concern with the
deformations of surfaces from integrable (2+1)-dimensional spin systems and
equations of the nonlinear Schr\"{o}dinger type, respectively. In Sections 9
and 10 some solutions and special surfaces associated with spin system
vortices are considered. Finally, in Section 11 some concluding remarks are
reported.

\bigskip

\section{ Chern-Simons theory and completely integrable systems}

Here we shall review some preliminaries concerning the CS theory and we show
how one can connect them to the completely integrable systems in 2+1
dimensions. We are dealing with the field theory defined by the action 
\begin{equation}
S\left[ J\right] =\frac{k}{4\pi }\int_{M}Tr\left( J\wedge dJ+\frac{2}{3}%
J\wedge J\wedge J\right) ,  \label{(1)}
\end{equation}
where $J$ is a 1-form gauge connection taking values in a simple Lie algebra 
$\hat{g}$ on an oriented closed three-manifold $M$, and $k$ is a coupling
constant which should be quantized in a quantum theory [9]. The related
classical equation of motion is the zero - curvature condition 
\begin{equation}
F\equiv dJ+J\wedge J=0.  \label{(2)}
\end{equation}

The action (\ref{(1)}) is manifestly invariant under general coordinate
transformations (preserving orientation and volumes). Moreover, under a
generic gauge map $G:{M}\rightarrow {\ }\widehat{{G}}$ the gauge connection
transforms as $J\rightarrow G^{-1}JG+G^{-1}dG$ . Correspondingly, the action
(\ref{(1)}) changes as $S[J]\rightarrow S[J]+2\pi \,k\,W(G),$ where 
\begin{equation}
W(J)={\frac{1}{24\pi ^{2}}}\int_{{M}}{Tr}(G^{-1}dG\wedge G^{-1}dG\wedge
G^{-1}dG)
\end{equation}
is the winding number of the map $G$ and takes integer values, because of
the result $\pi _{3}(\widehat{{G}})={\bf Z}$ from the homotopy theory [10].
This is a Topological Field Theory in the sense that it possesses quantum
observables, which are independent of the metric and are related to the
Jones polynomials of the knot theory [11]. From other points of view, the
action (\ref{(1)}) has been used as an effective interaction for
quasi-particles and vortices in two space dimensional systems, of interest
in the physics of high temperature superconductivity [12], and in the
context of the low dimensional gravity models (see [13] and reference
therein). In the static self-dual reductions the system of equations (\ref
{(2)}) becomes the two - dimensional Toda field theory [14], the static
reductions of the Ishimori model or of the Davey - Stewartson equation [15].

In [8] the general action (\ref{(1)}) was reduced assuming that the Lie
algebra $\hat{g}$ admits a ${\bf Z}_{2}$ graduation, in such a way the form $%
J$ splits in two parts, taking values on an isotropy subalgebra and a
complement linear space, respectively. The former component will play the
role of a gauge field with the isotropy group as a gauge group, the latter
could be considered as a sort of coupled ''matter'' field. At any point of
the corresponding coset space we can introduce in a natural way a Riemannian
torsion free connection [10]. Furthermore, the 3-manifold $M$ is trivialized
into $\Sigma \times {\bf R}$, where $\Sigma $ is a Riemann surface, endowed
with a set of local complex coordinates $z=x^{1}+ix^{2},$ and ${\bf R}$ is
interpreted as the time axis. Thus the connection $J$ can be further
decomposed into time and space-like components. In the simplest $SU\left(
2\right) /U(1)\sim CP^{1}\sim S^{2}$ case the connection takes the form 
\[
J=\left( 
\begin{array}{ll}
iv_{\mu } & -q_{\mu }^{\ast } \\ 
q_{\mu } & -iv_{\mu }
\end{array}
\right) dx^{\mu }, 
\]
which can be rewritten in terms of complex forms 
\[
vdz=\frac{1}{2}\left( v_{1}-iv_{2}\right) dz,\quad \psi _{-}dz=\frac{1}{2}%
\left( q_{1}-iq_{2}\right) dz,\quad \psi _{+}d\overline{z}=\frac{1}{2}\left(
q_{1}+iq_{2}\right) d\overline{z}. 
\]

\bigskip This allows us to write the action (\ref{(1)}) as follows

\[
S=-\frac{k}{\pi }\int_{\Sigma \times {\bf R}}\{\frac{1}{2}\varepsilon
^{\lambda \mu \nu }v_{\lambda }\partial _{\mu }v_{\nu }+\frac{i}{2}(\psi
_{+}^{\ast }D_{0}\psi _{+}-\psi _{+}(D_{0}\psi _{+})^{\ast }-\psi _{-}^{\ast
}D_{0}\psi _{-}+\psi _{-}(D_{0}\psi _{-})^{\ast }) 
\]

\begin{equation}
-iq_{0}^{*}(D\psi _{+}-\stackrel{\_}{D}\psi _{-})+iq_{0}(D\psi _{+}-%
\stackrel{\_}{D}\psi _{-})^{*}\}dx^{0}dx^{1}dx^{2},  \label{4}
\end{equation}

where $D_{0}=\partial _{0}-2iv_{0}$, $D=\partial _{z}-2iv$, $\overline{D}%
=\partial \overline{_{z}}-2iv^{\ast }$ ($^{\ast }$ denotes the complex
conjugation). The first order Lagrangian involved in (\ref{4}) is
constrained by the torsion-free condition 
\begin{equation}
{D}\psi _{+}-{\bar{D}}\psi _{-}=0  \label{torfree}
\end{equation}
and by what we call the Gauss - Chern - Simons (GCS) law 
\begin{equation}
{\partial }_{z}{v}^{\ast }-{\partial }_{\overline{z}}v=-i\left( {{\left| {{%
\psi }_{+}}\right| }^{2}-{\left| {{\psi }_{-}}\right| }^{2}}\right) ,
\label{GCS}
\end{equation}
enforced by the Lagrangian multipliers $q_{0}$ and $v_{0}$, respectively.%
{\bf \ }Of course, here we are looking in a different way to a subset of the
equations of motion (\ref{(2)}), in which the general covariance is broken.
Indeed, only the isotropic $U(1)$ invariance is left. Furthermore, by
exploiting the local isomorphism between $so(3)$ and $su(2)$ realized by the
adjoint representation of the connection 
\[
J^{(ad)}=\left( 
\begin{array}{lll}
0 & -v_{0} & -Re(q_{0}) \\ 
v_{0} & 0 & -Im(q_{0}) \\ 
Re(q_{0}) & Im(q_{0}) & 0
\end{array}
\right) dx^{0}+ 
\]
\[
\left( 
\begin{array}{lll}
0 & -2Re(v) & -Re(\psi _{+}+\psi _{-}) \\ 
2Re(v) & 0 & -Im(\psi _{+}+\psi _{-}) \\ 
Re(\psi _{+}+\psi _{-}) & Im(\psi _{+}+\psi _{-}) & 0
\end{array}
\right) dx^{1}+ 
\]
\[
\left( 
\begin{array}{lll}
0 & 2Im(v) & -Im(\psi _{+}-\psi _{-}) \\ 
-2Im(v) & 0 & -Re(\psi _{+}-\psi _{-}) \\ 
Im(\psi _{+}-\psi _{-}) & Re(\psi _{+}-\psi _{-}) & 0
\end{array}
\right) dx^{2}, 
\]
we are able to introduce the so-called moving trihedral frame $\{{\bf e}%
_{i}\}$ in ${\bf R}^{3}$ [10], which satisfies the orthonormal conditions 
\begin{equation}
{\bf e}_{i}\cdot {\bf e}_{j}=\delta _{ij},\,\,\,{\bf e}_{i}\wedge {\bf e}%
_{j}=\varepsilon _{ijk}{\bf e}_{k}
\end{equation}
and changes accordingly to 
\begin{equation}
\partial _{\mu }{\bf e}_{i}=(J_{\mu })_{ik}^{(ad)}{\bf e}_{k},\quad \mu
=0,1,2,\quad i=1,2,3.  \label{ntra}
\end{equation}
Its integrability is assured by the zero curvature condition, namely by Eq. (%
\ref{(2)}). For instance, assigning to ${\bf e}_{3}$ the special role of
unimodular normal vector to a given surface ${\cal S}$, whose tangent plane
is defined by the vectors $\left( {\bf e}_{1},{\bf e}_{2}\right) $, the
equations (\ref{ntra}) for $\mu =1,2$ and $i=1,2,3$ can be seen as the
Gauss-Weingarten equations of such a surface. Moreover, the mapping ${\bf e}%
_{3}:{\cal S}\longrightarrow S^{2}$ is the well known Gauss map.
Furthermore, the corresponding integrability equations, rewritten as 
\[
\partial _{1}J_{2}^{\left( ad\right) }-\partial _{2}J_{1}^{\left( ad\right)
}+\left[ J_{1}^{\left( ad\right) },J_{2}^{\left( ad\right) }\right] =0, 
\]
are the Gauss-Codazzi-Mainardi equations for a surface $S$ immersed in ${\bf %
R}^{3}.$ They are the real form of the equations (\ref{torfree}) and (\ref
{GCS}). The $U(1)$ invariance of such equations is readily interpreted as
the invariance under local rotations of the tangent plane at the surface $%
{\cal S}$. This identification is one of the motivation of the present
article and it will be fully developed from the geometrical point of view in
the next Sections. Here we would like to show how to use effectively the
remaining equations in (\ref{ntra}) for $\mu =0$ and $i=1,2,3$ and the
corresponding integrability conditions. In particular, we ask if structures
related to the integrable systems can be detected in the above general
picture. Then, since we show that it is the case, we are allowed to
introduce a completely integrable dynamics for the trihedral frame and, by
consequence, we can infer an integrable dynamics for the corresponding
surfaces.

First, we can rewrite Eqs. (\ref{torfree}) and (\ref{GCS}) by introducing
the quantities

\[
V=\left( 
\begin{array}{ll}
v* &  \\ 
& v
\end{array}
\right) ,\quad \Psi _{\pm }=\left( 
\begin{array}{ll}
& \psi _{\pm } \\ 
-\psi _{\pm }^{*} & 
\end{array}
\right) ,\quad B=\frac{i}{2}\sigma _{3}\left( \Psi _{-}-\Psi _{+}\right) . 
\]
Indeed, the torsionless condition (\ref{torfree}) and its complex conjugate
can be written as

\[
\left( 
\begin{array}{ll}
\partial _{z} &  \\ 
& \partial _{\bar{z}}
\end{array}
\right) B+\frac{i}{2}\left( \partial _{\bar{z}}-\partial _{z}\right) \Psi
_{-}+V\Psi _{-}-\Psi _{+}V=0, 
\]
while combining the GCS law (\ref{GCS}) and its complex conjugate yields 
\[
{\rm tr}\left\{ \sigma _{3}\left[ \left( 
\begin{array}{ll}
\partial _{z} &  \\ 
& \partial _{\bar{z}}
\end{array}
\right) V+B\Psi _{-}-\Psi _{+}B\right] \right\} =0. 
\]

These equations involve independent components in the basis of the complex $%
2\times 2$ matrices, and we have no information only about the identity
component $\sigma _{0}$. So the last two equations provide 
\begin{equation}
L_{+}\left[ \frac{i}{2}\left( \partial _{\bar{z}}-\partial _{z}\right) +V+B%
\right] -\left[ \frac{i}{2}\left( \partial _{\bar{z}}-\partial _{z}\right)
+V+B\right] L_{-}=f\sigma _{0},  \label{9}
\end{equation}

\begin{equation}
L_{\pm }=\left( 
\begin{array}{ll}
\partial _{z} &  \\ 
& \partial _{\bar{z}}
\end{array}
\right) -\Psi _{\pm }
\end{equation}
has the form of the two-dimensional principal Zakharov-Shabat spectral
problem [16] of elliptic type. Moreover, putting $f\equiv 0$ in equation (%
\ref{9}), we obtain the so called space part of the B\"{a}cklund
transformations associated with $L_{\pm }$, defining the first-order B\"{a}%
cklund-gauge operator 
\[
\hat{B}=\frac{i}{2}\left( \partial _{\bar{z}}-\partial _{z}\right) +V+B.
\]
This means that $\hat{B}$ maps solutions of the two linear problems 
\begin{equation}
L_{-}\phi _{-}=0\Rightarrow L_{+}B\phi _{-}=L_{+}\phi _{+}=0.
\label{spectralProb}
\end{equation}

The previous considerations tell us that in the above formalism the Gauss -
Codazzi - Mainardi equations are expressed in terms of products of first
order differential operators, which have a precise meaning in the theory of
the completely integrable systems. Of course, in this context an essential
role is played by the second (evolution) linear operator of the Lax pair, in
order to introduce a compatible time evolution. The latter, generally
speaking, is non linear, while the corresponding equations from the CS
theory still contains the arbitrary functions interpreted as Lagrangian
multipliers in the action (\ref{4}). Indeed, the corresponding equations
read 
\begin{equation}
D_{0}\psi _{+}=\bar{D}~q_{0},\qquad D_{0}\psi _{-}=D~q_{0},  \label{PreDS}
\end{equation}
\begin{equation}
\partial _{0}v-\partial _{z}v_{0}=i\left( q_{0}\psi _{+}^{\ast }-q_{0}^{\ast
}\psi _{-}\right)  \label{electric}
\end{equation}
and their complex conjugated. However, we can exploit the freedom in the
choice of $q_{0}$ and $v_{0}$ in order to fix the evolution of $\psi _{\pm }$
and $v$ in the $x^{0}$ variable. In fact, let us take 
\begin{equation}
q_{0}=2i\left[ \left( \bar{D}-\frac{i}{2}\left( \partial _{\bar{z}}\omega
-i\partial _{\bar{z}}\chi \right) \right) \psi _{+}+\left( D-\frac{i}{2}%
\left( \partial _{z}\omega +i\partial _{z}\chi \right) \right) \psi _{-}%
\right] ,  \label{qO}
\end{equation}
where we require that the real functions $\chi $ and $\omega $ satisfy the
supplementary conditions 
\begin{equation}
\partial _{z}\partial _{\bar{z}}\chi =-4\left( {{\left| {{\psi }_{+}}\right| 
}^{2}-{\left| {{\psi }_{-}}\right| }^{2}}\right) ,\quad \partial
_{z}\partial _{\bar{z}}\omega =0.  \label{qO+}
\end{equation}
Furthermore, by introducing an irrotational field ${\bf A}=\left( \partial
_{z}\Lambda ,\partial _{\bar{z}}\Lambda ,\partial _{0}\Lambda \right) $
 ( $\Lambda$  is an arbitrary real function)  in such a way
that

\[
v=\frac{1}{4}\partial _{z}\Lambda -\frac{i}{8}\partial _{z}\chi ,\quad v_{0}=%
\frac{1}{4}\left( \partial _{0}\Lambda +u_{0}\right) ,\quad \psi _{\pm
}=\Psi _{\pm }\exp \left( \frac{i}{2}\Lambda \right) , 
\]
the ''time evolutions'' (\ref{PreDS}) become

\begin{equation}
i\partial _{0}\Psi _{\pm }+2\left( \partial _{z}^{2}+\partial _{\bar{z}%
}^{2}\right) \Psi _{\pm }+\frac{1}{2}u_{0}^{\pm }\Psi _{\pm }-i\left(
\partial _{\bar{z}}\omega \partial _{\bar{z}}+\partial _{z}\omega \partial
_{z}\right) \Psi _{\pm }=0,  \label{DSlike}
\end{equation}
where we have suitably defined in terms of $u_{0,}\chi $ and $\omega $ the
scalar fields $u_{0}^{\pm }$ , which obey the consistency conditions arising
from (\ref{electric}) 
\begin{equation}
\partial _{z}\partial _{\bar{z}}u_{0}^{\pm }=8\left( \partial
_{z}^{2}+\partial _{\bar{z}}^{2}\right) \left| \Psi _{\pm }\right| ^{2}.
\label{DSscalar}
\end{equation}
Thus, to summarize, the gauge fixing conditions (\ref{qO}) and (\ref{qO+})
destroy the arbitraryness contained in the equations (\ref{PreDS}) in favour
of a formally decoupled pair of Davey - Stewartson equations (\ref{DSlike})
and (\ref{DSscalar}). Actually, between the two pairs of fields $\left( \Psi
_{\pm },u_{0}^{\pm }\right) $ there still exists the coupling provided by
the torsionless condition (\ref{torfree}), which in the new variables takes
the form 
\begin{equation}
\left( \partial _{\bar{z}}+\frac{1}{4}\partial _{\bar{z}}\chi \right) \Psi
_{-}=\left( \partial _{z}+\frac{1}{4}\partial _{z}\chi \right) \Psi _{+}.
\label{pseudoBT}
\end{equation}
As we discussed above, Eq. (18) is in essence the space part of the
B\"{a}cklund transformations. Starting from a known solution, say $\left(
\Psi _{-},u_{0}^{-}\right) $ and fixing $\omega ,$ one can reconstruct the
function $\chi $, and solving (\ref{pseudoBT}) for $\Psi _{+},$ finally we
find $u_{0}^{+}$ from (\ref{DSscalar}). Furthermore, we observe that the
gauge choice (\ref{qO}) - (\ref{qO+}) is equivalent to fix the second
operator of the Lax pair, denoted here by $M_{\pm }=\partial _{x^{0}}+\Sigma
_{k=0}^{2}M_{\pm }^{\left( k\right) }\left[ \frac{i}{2}\left( \partial _{%
\bar{z}}-\partial _{z}\right) \right] ^{2-k}$, where $M_{\pm }^{\left(
k\right) }$ are specific matrices. $M_{\pm }$ provide the system (\ref
{DSlike}) - (\ref{DSscalar}) and (\ref{pseudoBT}) by the compatibility
relations 
\[
\left[ L_{\pm },M_{\pm }\right] =0,\quad M_{+}\hat{B}-\hat{B}M_{-}=0. 
\]

Moreover, by using a suitable particular eigenfunction $\phi _{-}^{0}$ of
the equation (\ref{spectralProb}), it is well known (see [17],[18]) that one
can construct a new spectral problem of the form

\[
L_{-}^{I}=\left( \phi _{-}^{0}\right) ^{-1}L_{-}\phi _{-}^{0}=\partial
_{x^{1}}+iS\partial _{x^{2}},\quad M_{-}^{I}=\left( \phi _{-}^{0}\right)
^{-1}M_{-}\phi _{-}^{0}, 
\]
where 
\[
S=i\left( \phi _{-}^{0}\right) ^{-1}\sigma _{3}\phi _{-}^{0} 
\]
is an element of $SU\left( 2\right) /U\left( 1\right) $ coset space and the
corresponding eigenfunction is $\phi _{-}^{I}=\left( \phi _{-}^{0}\right)
^{-1}\phi _{-}.$ The resulting integrable system is known as Ishimori model
and it describes the evolution of a classical spin in a background generated
by the density of the topological charge. Since this equation will be
discussed in the next Sections, we do not give other details about it. But
here we want to stress that such a system is an alternative integrable
restriction of the possible configurations of the CS field, exactly as the
Davey-Stewartson equation does. Furthermore, one can put the question if the
spin field $S$ has something to do with the trihedral frame introduced
above. The consequences arising from the identification of $S$ with one of
the unimodular vector fields ${\bf e}_{i}$ is the main subject of the next
Sections.

\bigskip

\section{Surfaces in $R^{3}$}

To introduce our terminology and notations and to make the exposition
self-contained, we recall some basic facts from the theory of 2-dimensional
surfaces. So, we consider a smooth surface in a three dimensional Euclidean
space{\bf \ }${\bf R}^{3}$. Let $x,y$ be the local coordinates on the
surface. At the same time, the surface can be described by the position
vector ${\bf r}(x_{1},x_{2},x_{3})={\bf r}(x,y)$, where $x_{i}$ are
coordinates of ${\bf R}^{3}$. The surface is uniquely defined within rigid
motions by the two fundamental forms \newline
\begin{equation}
I=Edx^{2}+2Fdxdy+Gdy^{2}
\end{equation}

\bigskip\ and 
\begin{equation}
II=Ldx^{2}+2Mdxdy+Ndy^{2},
\end{equation}

where $E,F,G,L,M,N$ can be defined by

\begin{equation}
E={\bf r}_{x}\cdot {\bf r}_{x}=g_{11},\quad F={\bf r}_{x}\cdot {\bf r}%
_{y}=g_{12}=g_{21},\quad G={\bf r}_{y}\cdot {\bf r}_{y}=g_{22},
\end{equation}

\bigskip

\begin{equation}
L={\bf n\cdot }{\bf r}_{xx}=b_{11},\quad M={\bf n\cdot }{\bf r}%
_{xy}=b_{12}=b_{21},\quad N={\bf n\cdot }{\bf r}_{yy}=b_{22}.
\end{equation}

\bigskip

In Equations (21) and (22) 
\begin{equation}
{\bf n}(x,y)=\frac{{\bf r}_{x}\wedge {\bf r}_{y}}{\mid {\bf r}_{x}\wedge 
{\bf r}_{y}\mid }=\frac{{\bf r}_{x}\wedge {\bf r}_{y}}{\sqrt{g}}\quad
\end{equation}

is introduced, where $g=det(g_{ij})=EG-F^{2}=\mid {\bf r}_{x}\wedge {\bf r}%
_{y}\mid ^{2},$ is the normal vector field at each point of the surface.
Then the triple $({\bf r}_{x},{\bf r}_{y},{\bf n})$ represents a local frame
of ${\bf R}^{3},$ the changes of which are characterized by the GW equations 
\begin{equation}
{\bf r}_{xx}=\Gamma _{11}^{1}{\bf r}_{x}+\Gamma _{11}^{2}{\bf r}_{y}+L{\bf n,%
}
\end{equation}

\begin{equation}
{\bf r}_{xy}=\Gamma _{12}^{1}{\bf r}_{x}+\Gamma _{12}^{2}{\bf r}_{y}+M{\bf n}%
,
\end{equation}

\begin{equation}
{\bf r}_{yy}=\Gamma _{22}^{1}{\bf r}_{x}+\Gamma _{22}^{2}{\bf r}_{y}+N{\bf n,%
}
\end{equation}

\begin{equation}
{\bf n}_{x}=p_{11}{\bf r}_{x}+p_{12}{\bf r}_{y},
\end{equation}

\begin{equation}
{\bf n}_{y}=p_{21}{\bf r}_{x}+p_{22}{\bf r}_{y},
\end{equation}

where the Christoffel symbols of the second kind are defined by $g_{ij}$ ( $%
g^{ij}=(g_{ij})^{-1})$ as 
\begin{equation}
\Gamma _{jk}^{i}=\frac{1}{2}g^{il}(\frac{\partial g_{lk}}{\partial x^{j}}+%
\frac{\partial g_{jl}}{\partial x^{k}}-\frac{\partial g_{jk}}{\partial x^{l}}%
),
\end{equation}

and 
\begin{equation}
p_{ij}=-b_{ik}g^{kj}.
\end{equation}

The principal curvatures $k_{1},k_{2}$ are the eigenvalues of the Weingarten
operator 
\begin{equation}
\Lambda =\left( 
\begin{array}{cc}
E & F \\ 
F & G
\end{array}
\right) ^{-1}\left( 
\begin{array}{cc}
L & M \\ 
M & N
\end{array}
\right) .
\end{equation}

which for the mean and the Gassian curvature implies 
\begin{equation}
H=\frac{k_{1}+k_{2}}{2}=tr(\Lambda )=\frac{EN+LG-2MF}{2(EG-F^{2}},
\end{equation}

\begin{equation}
K=k_{1}k_{2}=det(\Lambda )=\frac{LN-M^{2}}{EG-F^{2}}.
\end{equation}

One of the global characteristics of surfaces is the integral curvature 
\begin{equation}
\chi =\frac{1}{2\pi }\int K\sqrt{g}dxdy,
\end{equation}

which for compact oriented surfaces is the integer 
\begin{equation}
\chi =2(1-\triangle ),
\end{equation}

where $\triangle $ is the genus of the surface. The compatibility conditions
of the GW equations (24)-(28) furnish the GMC equations 
\begin{equation}
R_{ijk}^{l}=b_{ij}b_{k}^{l},\quad \frac{\partial b^{ij}}{\partial x^{k}}-%
\frac{\partial b_{ik}}{\partial x^{j}}=\Gamma _{ik}^{s}b_{js}-\Gamma
_{ij}^{s}b_{ks},
\end{equation}

where $b_{i}^{j}=g^{jl}b_{il}$ and the curvature tensor is 
\begin{equation}
R_{ijk}^{l}=\frac{\partial \Gamma _{ij}^{l}}{\partial x^{k}}-\frac{\partial
\Gamma _{ik}^{l}}{\partial x^{j}}+\Gamma _{ij}^{s}\Gamma _{ks}^{l}-\Gamma
_{ik}^{s}\Gamma _{js}^{l}.
\end{equation}

For our purposes it is convenient to employ the triad of orthonormal vectors 
\begin{equation}
{\bf e}_{1}=\frac{{\bf r}_{x}}{\sqrt{E}},\quad {\bf e}_{2}={\bf n},\quad 
{\bf e}_{3}={\bf e}_{1}\wedge {\bf e}_{2}.
\end{equation}
In terms of these vectors the GW equations (24)-(28) take the form 
\begin{equation}
{\bf e}_{jx}={\bf X}\wedge {\bf e}_{j},\quad {\bf e}_{jy}={\bf Y}\wedge {\bf %
e}_{j},
\end{equation}
where 
\begin{equation}
{\bf X}=\tau {\bf e}_{1}+\sigma {\bf e}_{2}+k{\bf e}_{3},\quad {\bf Y}=m_{1}%
{\bf e}_{1}+m_{2}{\bf e}_{2}+m_{3}{\bf e}_{3},
\end{equation}
and 
\begin{equation}
k=\frac{L}{\sqrt{E}},\quad \sigma =\frac{2EF_{x}-EE_{y}-FE_{x}}{2E\sqrt{g}}%
,\quad \tau =\frac{ME-LF}{\sqrt{gE}},
\end{equation}
\begin{equation}
m_{1}=\frac{NE-MF}{\sqrt{gE}},\quad m_{2}=\frac{FE_{y}-EG_{x}}{2E\sqrt{g}}%
,\quad m_{3}=\frac{M}{\sqrt{E}}.
\end{equation}

Similarly, we can rewrite the GMC equations (32) in the following form 
\begin{equation}
A_{y}-B_{x}+[A,B]=0,
\end{equation}
with 
\begin{equation}
A=\left( 
\begin{array}{ccc}
0 & k & -\sigma \\ 
-k & 0 & \tau \\ 
\sigma & -\tau & 0
\end{array}
\right) ,\quad B=\left( 
\begin{array}{ccc}
0 & m_{3} & -m_{2} \\ 
-m_{3} & 0 & m_{1} \\ 
m_{2} & -m_{1} & 0
\end{array}
\right) .
\end{equation}

Then, the GMC equation turns out to be equivalent to the set of equations
for the coefficients of the first and second fundamental forms. This system,
which is in general non integrable, reduces to integrable partial
differential equations for certain particular surfaces [10].

\bigskip

\section{Deformations of surfaces in 2+1 dimensions: the general formulation}

\bigskip

It is well known that in some cases deformations of surfaces can be
associated with integrable equations [1-3]. Here we are interested in the
deformation of the two-dimensional surfaces discussed in Section 3. In other
words, we have to deal with the motion of such surfaces. To this aim, let us
introduce the vector field

\begin{equation}
{\bf r}_{t}=a_{1}{\bf r}_{x}+a_{2}{\bf r}_{y}+a_{3}{\bf n},
\end{equation}
where the $a_{i}$'s are some real functions. It is easy to show that the
evolution of the local trihedral frame is given by 
\begin{equation}
{\bf e}_{jt}={\bf T}\wedge {\bf e}_{j},
\end{equation}
\begin{equation}
{\bf T}=\omega _{1}{\bf e}_{1}+\omega _{2}{\bf e}_{2}+\omega _{3}{\bf e}_{3},
\end{equation}
$\omega _{j}$'s being real functions. Summarizing, the changes of the local
frame are provided by

\begin{equation}
{\bf e}_{jx}={\bf X}\wedge {\bf e}_{j},\quad {\bf e}_{jy}={\bf Y}\wedge {\bf %
e}_{j},\quad {\bf e}_{jt}={\bf T}\wedge {\bf e}_{j},
\end{equation}

where the vectors{\bf \ X }and {\bf Y }are defined by (40). This system is
analogous to the system (8) in Section 2.

The system (52) represents the simplest form of the (2+1)-dimensional GW
equations.

By introducing the matrix

\begin{equation}
C=\left( 
\begin{array}{ccc}
0 & \omega _{3} & -\omega _{2} \\ 
-\omega _{3} & 0 & \omega _{1} \\ 
\omega _{2} & -\omega _{1} & 0
\end{array}
\right)
\end{equation}
and using the matrices $A$ and $B$ (see (44)), the compatibility conditions
of Eqs. (48) entail 
\begin{equation}
A_{y}-B_{x}+[A,B]=0,
\end{equation}
\begin{equation}
A_{t}-C_{x}+[A,C]=0,
\end{equation}
\begin{equation}
B_{t}-C_{y}+[B,C]=0.
\end{equation}

Of the nine funcions $k,\sigma ,\tau ,m_{i},\omega _{i}$ involved in $A,B$
and $C$ only three are independent. In fact, we can express the functions $%
m_{i},\omega _{i}$ in terms of $k,\sigma ,\tau $ and their derivatives. This
point will be discussed later.

\subsection{Some geometrical invariants and integrals of motion as
consequence of the geometrical formalism}

The formalism developed above yields some important invariants having a pure
geometrical nature. Indeed, in terms of the triad vectors these geometrical
invariants take the form 
\[
K_{1}^{(t)}=\int {\bf e}_{1}\cdot ({\bf e}_{1x}\wedge {\bf e}%
_{1y})dxdy,\quad K_{2}^{(t)}=\int {\bf e}_{2}\cdot ({\bf e}_{2x}\wedge {\bf e%
}_{2y})dxdy,\quad 
\]

\begin{equation}
K_{3}^{(t)}=\int {\bf e}_{3}\cdot ({\bf e}_{3x}\wedge {\bf e}_{3y})dxdy.
\end{equation}

In a similar way we can write down other two classes of invariants with
respect to $x$ and $y$ directions, respectively. These geometrical
invariants can be interpreted as ''topological charges''. However, three of
them, namely $K_{i}^{(t)}$ $(i=1,2,3)$ behave as integrals of motion of the
(2+1)-dimensional geometrical models under consideration. This will be
elucidated in the next Sections. These invariants can be related to the
topological Chern index of a curvature 2-form on a 2-dimensional space [19].

\section{Integrable spin models in (2+1)- dimensions}

Now let us dwell upon the problem of finding or building up integrable
deformations of (2+1)-dimensional surfaces. Among several possibilities,
within the geometrical formalism previously presented we shall consider
multidimensional integrable spin (field) systems (MISSs) to recognize
integrable deformations of surfaces.

\subsection{The spin model}

A few words on MISSs. At present there exist many integrable spin systems in
(2+1)-dimensions (see, for example, Refs. [20,21-25]). A well known
prototype of these systems is the Ishimori model (IM) [20]. A more general
(2+1)-dimensional integrable spin model is described by the pair of equations

\begin{equation}
{\bf S}_{t}+{\bf S}\wedge \{(b+1){\bf S}_{\xi \xi }-b{\bf S}_{\eta \eta
}\}+bu_{\eta }{\bf S}_{\eta }+(b+1)u_{\xi }{\bf S}_{\xi }=0,
\end{equation}
\begin{equation}
u_{\xi \eta }={\bf S\cdot }({\bf S}_{\xi }\wedge {\bf S}_{\eta }),
\end{equation}
where $\xi ,\eta $ are real or complex variables, $b$ is a real constant, $%
{\bf S}=(S_{1},S_{2},S_{3})$ is the spin (field) vector, ${\bf S}^{2}=1$,
and $u$ is a scalar function. These equations, which are called M-XX
equations (about our conditional notations, see e.g. [21-25]), are one of
the (2+1)-dimensional integrable generalizations of the isotropic
Landau-Lifshitz (LL) equation 
\begin{equation}
{\bf S}_{t}={\bf S}\wedge {\bf S}_{xx}.
\end{equation}
In (1+1)-dimensions, Eqs.(54) and (55) reduce to the LL equation. In fact,
assuming that the variables ${\bf S},u$ are, for example, independent of $%
\eta ,$ then Eqs. (54) and (55) reproduce Eq. (56) within a simple scale
transformation.

We notice that Eqs. (54) and (55) are not the only integrable generalization
of the LL equation in (2+1)-dimensions. Actually, other integrable
generalizations exist, such as the IM or the model defined by 
\begin{equation}
{\bf S}_{t}=({\bf S}\wedge {\bf S}_{y}+u{\bf S})_{x},
\end{equation}
\begin{equation}
u_{x}=-{\bf S\cdot }({\bf S}_{x}\wedge {\bf S}_{y}).
\end{equation}
These equations, which are called M-I equations (see [21]) are again
completely integrable. Some properties of these equations are studied in
[21-23].

\subsubsection{The Lax representation}

Equations (54) and (55) can be solved by the IST method. The applicability
of the IST method to Eqs. (54) and (55) is based on the equivalence of these
equations to the compatibility condition of the following linear equations
(the Lax representation (LR)) 
\begin{equation}
\Phi _{Z^{+}}=S\Phi _{Z^{-}},
\end{equation}
\begin{equation}
\Phi _{t}=2i[S+(2b+1)I]\Phi _{Z^{-}Z^{-}}+W\Phi _{Z^{-}},
\end{equation}
where $Z^{\pm }=\xi \pm \eta $ and 
\[
W=2i\{(2b+1)(F^{+}+F^{-}S)+(F^{+}S+F^{-})+(2b+1)SS_{Z^{-}}+\frac{1}{2}%
S_{Z^{-}}+\frac{1}{2}SS_{Z^{+}}\},\quad 
\]
\[
S=\pmatrix{
S_3 & rS^- \cr
rS^+ & -S_3
},\quad S^{\pm }=S_{1}\pm iS_{2}\quad S^{2}=EI,\quad E=\pm 1,\quad r^{2}=\pm
1, 
\]
\[
F^{+}=2iu_{Z^{-}},\quad F^{-}=2iu_{Z^{+}}. 
\]
In fact, from the condition $\Phi _{Z^{+}t}=\Phi _{tZ^{+}}$ we deduce 
\begin{equation}
iS_{t}+\frac{1}{2}[S,(b+1)S_{\xi \xi }-bS_{\eta \eta }]+ibu_{\eta }S_{\eta
}+i(b+1)u_{\xi }S_{\xi }=0,
\end{equation}
\begin{equation}
u_{\xi \eta }=\frac{1}{4i}tr(S[S_{\xi },S_{\eta }]),
\end{equation}
which is the matrix form of Eqs. (54) and (55).

\subsubsection{Special cases}

Equations (54) and (55) contain both well known and less known integrable
cases in (2+1) and (1+1)-dimensions. Below we shall report some of them.

\bigskip

$i)$ If $b=0$, Eqs. (54) and (55) yield 
\begin{equation}
{\bf S}_{t}+{\bf S}\wedge {\bf S}_{\xi \xi }+w{\bf S}_{\xi }=0,
\end{equation}
\begin{equation}
w_{\eta }-{\bf S\cdot }({\bf S}_{\xi }\wedge {\bf S}_{\eta })=0,
\end{equation}
where $w=u_{\xi }$. This system, which is known as the M-VIII model [24], is
one of the simplest spin systems in (2+1)-dimensions integrable by IST. It
affords different type of solutions (solitons, vortices, etc.). In
particular, vortex solutions of Eqs. (54)-(55) can be derived from vortex
solutions of the spin system (54)-(55) discussed in Section 9 (for $b=0).$

\bigskip

$ii)$ Let us introduce the coordinates $x=\xi -\eta ,\quad y=\alpha (\xi
+\eta )$ and put $b=-\frac{1}{2}$. Then, the spin system (54)-(55) reduces
to the IM 
\begin{equation}
{\bf S}_{t}+{\bf S}\wedge ({\bf S}_{xx}+\alpha ^{2}{\bf S}_{yy})+u_{x}{\bf S}%
_{y}+u_{y}{\bf S}_{x}=0,
\end{equation}
\begin{equation}
u_{xx}-\alpha ^{2}u_{yy}=-2\alpha ^{2}{\bf S\cdot }({\bf S}_{x}\wedge {\bf S}%
_{y}).
\end{equation}

The IM is the first integrable spin (field) system in the plane which can be
solved by IST method. The IM was studied by many authors from different
points of view (e.g. [16,17,20,23,32]).

\bigskip

$iii)$ By setting $b=0,\eta =t$, Eqs. (54) and (55) reduce to the following
(1+1)-dimensional spin system: 
\begin{equation}
{\bf S}_{t}+{\bf S}\wedge {\bf S}_{\xi \xi }+w{\bf S}_{\xi }=0,
\end{equation}
\begin{equation}
w_{t}+\frac{1}{2}({\bf S}_{\xi }^{2})_{\xi }=0.
\end{equation}
This integrable model describes the nonlinear dynamics of compressible
magnets [26]. It is the first (and, to the best of our knowledge, at present
the unique) example of integrable spin system governing the nonlinear
interactions of spin (${\bf S}$) and lattice ($u$) subsystems in
(1+1)-dimensions.

\bigskip

\section{Bilinear representations}

One of the powerful tools in the soliton theory is the Hirota method. Now we
show how to construct the bilinear representations of the fields of the spin
system by using geometry. Let $e_{ji}$ are the components of the unit vector 
${\bf e}_{j}$, i.e. ${\bf e}_{j}=(e_{j1},e_{j2},e_{j3})$.We can take the
following representation for the components of the vector ${\bf e}_{1}:$ 
\begin{equation}
e_{1}^{+}=e_{11}+ie_{12}=\frac{2\bar{f}g}{\bar{f}f+\bar{g}g},\quad e_{13}=%
\frac{\bar{f}f-\bar{g}g}{\bar{f}f+\bar{g}g}.
\end{equation}
Then, we get 
\begin{equation}
e_{2}^{+}=e_{21}+ie_{22}=\frac{\bar{f}^{2}+g^{2}}{\bar{f}f+\bar{g}g},\quad
e_{23}=i\frac{fg-\bar{f}\bar{g}}{\bar{f}f+\bar{g}g},
\end{equation}
\begin{equation}
e_{3}^{+}=e_{31}+ie_{32}=\frac{\bar{f}^{2}-g^{2}}{\bar{f}f+\bar{g}g},\quad
e_{33}=-\frac{fg+\bar{f}\bar{g}}{\bar{f}f+\bar{g}g},
\end{equation}
with 
\begin{equation}
k=-i\frac{D_{x}(g\circ f-\bar{g}\circ \bar{f})}{\bar{f}f+\bar{g}g},\quad
m_{3}=-i\frac{D_{y}(g\circ f-\bar{g}\circ \bar{f})}{\bar{f}f+\bar{g}g},
\end{equation}
\begin{equation}
\sigma =-\frac{D_{x}(g\circ f+\bar{g}\circ \bar{f})}{\bar{f}f+\bar{g}g}%
,\quad m_{2}=-\frac{D_{y}(g\circ f+\bar{g}\circ \bar{f})}{\bar{f}f+\bar{g}g},
\end{equation}
\begin{equation}
\tau =i\frac{D_{x}(\bar{f}\circ f+\bar{g}\circ g)}{\bar{f}f+\bar{g}g},\quad
m_{1}=i\frac{D_{y}(\bar{f}\circ f+\bar{g}\circ g)}{\bar{f}f+\bar{g}g},
\end{equation}
\begin{equation}
\omega _{3}=-i\frac{D_{t}(g\circ f-\bar{g}\circ \bar{f})}{\bar{f}f+\bar{g}g}%
,\quad \omega _{2}=-\frac{D_{t}(g\circ f+\bar{g}\circ \bar{f})}{\bar{f}f+%
\bar{g}g}
\end{equation}
\begin{equation}
\omega _{1}=i\frac{D_{t}(\bar{f}\circ f+\bar{g}\circ g)}{\bar{f}f+\bar{g}g}.
\end{equation}
The Hirota operators $D_{x},D_{y}$ and $D_{t}$ are defined by 
\begin{eqnarray*}
D_{x}^{l}D_{y}^{m}D_{t}^{n}f(x,y,t)\circ g(x,y,t) &=&(\partial _{x}-\partial
_{x^{\prime }})^{l}(\partial _{y}-\partial _{y^{\prime }})^{m}(\partial
_{t}-\partial _{t^{\prime }})^{n}f(x,y,t)\circ \\
g(x^{\prime },y^{\prime },t^{\prime }) &\mid &_{x=x^{\prime },y=y^{\prime
},t=t^{\prime }}.
\end{eqnarray*}

Now we write down the bilinear representation for the spin vector and for
the derivatives of the potential $u.$ Taking into account (69)-(76), we find 
\begin{equation}
S^{+}=S_{1}+iS_{2}=\frac{2\bar{f}g}{\bar{f}f+\bar{g}g},\quad S_{3}=\frac{%
\bar{f}f-\bar{g}g}{\bar{f}f+\bar{g}g}.
\end{equation}
This is the general representation for the components of the spin vector for
all the spin systems. However, for the potential the bilinear forms for
every spin system should be different. In the following we shall consider
some examples.

\bigskip

$i)$ {\it The Ishimori model}. In this case we have 
\begin{equation}
\tau =\frac{1}{2}u_{y},\quad m_{1}=\frac{1}{2\alpha ^{2}}u_{x}.
\end{equation}
Hence, from (74) we get 
\begin{equation}
u_{y}=-2i\frac{D_{x}(\bar{f}\circ f+\bar{g}\circ g)}{\bar{f}f+\bar{g}g}%
,\quad u_{x}=-2i\alpha ^{2}\frac{D_{y}(\bar{f}\circ f+\bar{g}\circ g)}{\bar{f%
}f+\bar{g}g}.
\end{equation}

On the other hand, from (74) it follows also 
\begin{equation}
\tau _{x}=\alpha ^{2}m_{1y},
\end{equation}
so that 
\begin{equation}
m_{1}=\alpha ^{-2}\partial _{y}^{-1}\tau _{x}.
\end{equation}

\bigskip

$ii)$ {\it The isotropic M-I equation}. Let us take 
\begin{equation}
\tau =0,\quad m_{1}=u.
\end{equation}
Then, from (74) and (82) we obtain 
\begin{equation}
D_{x}(\bar{f}\circ f+\bar{g}\circ g)=0,\quad u=-i\frac{D_{y}(\bar{f}\circ f+%
\bar{g}\circ g)}{\bar{f}f+\bar{g}g}.
\end{equation}

$iii)${\it The spin system} (54)-(55). Let us start from 
\begin{equation}
\tau =\frac{1}{2}u_{\xi },\quad m_{1}=-\frac{1}{2}u_{\eta }.
\end{equation}
Then we have 
\[
u_{\xi }=-2i\frac{D_{\xi }(\bar{f}\circ f+\bar{g}\circ g)}{\bar{f}f+\bar{g}g}%
,\quad u_{\eta }=2i\frac{D_{\eta }(\bar{f}\circ f+\bar{g}\circ g)}{\bar{f}f+%
\bar{g}g}, 
\]
and 
\begin{equation}
\tau _{\eta }=-m_{1\xi },
\end{equation}
\begin{equation}
m_{1}=-\partial _{\xi }^{-1}\tau _{\eta }.
\end{equation}
An important consequence of these results is the possibility to determine
the time evolution of the potential (and/or its derivatives). For instance,
for the IM the time evolution of the derivatives of the potential are given
by 
\begin{equation}
\frac{1}{2}(u_{y})_{t}-\omega _{3x}+\sigma \omega _{1}-\tau \omega _{2}=0,
\end{equation}
\begin{equation}
\frac{1}{2\alpha ^{2}}(u_{x})_{t}-\omega _{1y}+m_{3}\omega _{3}-m_{2}\omega
_{3}=0.
\end{equation}

\section{Deformations of surfaces by integrable spin systems in 2+1
dimensions}

An interesting example of surface integrable deformation can be found out by
identifying the tangent unit vector ${\bf e}_{1}$ with the spin vector, i.e. 
\begin{equation}
{\bf e}_{1}\equiv {\bf S}.
\end{equation}
In such a way, the spin model (54)-(55) takes the form 
\begin{equation}
{\bf e}_{1t}+{\bf e}_{1}\wedge \{(b+1){\bf e}_{1\xi \xi }-b{\bf e}_{1\eta
\eta }\}+bu_{\eta }{\bf e}_{1\eta }+(b+1)u_{\xi }{\bf e}_{1\xi }=0,
\end{equation}
\begin{equation}
u_{\xi \eta }={\bf e}_{1}\cdot ({\bf e}_{1\xi }\wedge {\bf e}_{1\eta }).
\end{equation}
The functions $m_{i},\omega _{i}$ can be expressed in terms of the three
independent functions $k,\tau $,$\sigma .$Using the GW equations (48), Eqs.
(90)-(91) can be written as 
\begin{equation}
{\bf e}_{1t}=\omega _{3}{\bf e}_{2}-\omega _{2}{\bf e}_{3},
\end{equation}
\begin{equation}
u_{\xi \eta }=\sigma m_{3}-km_{2},
\end{equation}
where 
\begin{equation}
\omega _{2}=b[m_{3\eta }-m_{2}^{2}-u_{\eta }m_{2}]-(b+1)[k_{\xi }+\sigma
\tau +u_{\xi }\sigma ],
\end{equation}
\begin{equation}
\omega _{3}=(b+1)[\sigma _{\xi }-k\tau -ku_{\xi }]-b[m_{2\eta
}-m_{1}m_{3}+u_{\eta }m_{3}].
\end{equation}

\bigskip

Now by choosing $m_{1}$ according to the special reduction (84), the
remaining functions $m_{2}$ and $m_{3}$ are given by 
\begin{equation}
\quad m_{2}=\frac{\sigma m_{3}-u_{\xi \eta }}{k},\quad m_{3}=\frac{\sigma
\tau u_{\xi \eta }+(a_{2}/2)u_{\xi }-a_{3}}{a_{1}+\sigma ^{2}\tau },
\end{equation}

\bigskip where

\[
a_{1}=k^{2}\tau -k\sigma _{\xi }+\sigma k_{\xi },a_{2}=-k^{3}-\sigma ^{2}k, 
\]

\begin{equation}
a_{3}=k^{2}\sigma _{\eta }-\sigma kk_{\eta }+ku_{\xi \xi \eta }-k_{\xi
}u_{\xi \eta }.
\end{equation}

By virtue of these formulae we derive the function $\omega _{1}$ from (50).

Thus, all the unknown functions $m_{i},\omega _{i}$ are defined via the
three functions $k,\tau ,\sigma $ only and their derivatives. This is the
consequence of the identification of the motion of surface with the spin
system (90)-(91).This means that the motion of surface is fully determined
by these three functions. Since the spin model (90)-(91) is integrable, we
can conclude that the deformation of the surface characterized by Eqs.
(50)-(52) is integrable.

\section{Deformations of surfaces related to the (2+1)-dimensional NLS-type
equation}

One of the most remarkable consequence of the geometrical formalism
previously outlined is that it allows to find the equivalent counterpart of
the spin system (54)-(55). To show this property, let us introduce two
complex functions $q,p$ according to the following expressions 
\begin{equation}
q=a_{1}e^{ib_{1}},\quad p=a_{2}e^{ib_{2}},
\end{equation}
where $a_{j},b_{j}$ are real functions. Now let us choose the functions $%
a_{j},b_{j}$ in such a way that 
\begin{equation}
a_{1}^{2}=\frac{1}{4}k^{2}+\frac{|\alpha |^{2}}{4}(m_{3}^{2}+m_{2}^{2})-%
\frac{1}{2}\alpha _{R}km_{3}-\frac{1}{2}\alpha _{I}km_{2},
\end{equation}
\begin{equation}
b_{1}=\partial _{x}^{-1}\{-\frac{\gamma _{1}}{2ia_{1}^{\prime ^{2}}}-(\bar{A}%
-A+D-\bar{D})\},
\end{equation}
\begin{equation}
a_{2}^{2}=\frac{1}{4}k^{2}+\frac{|\alpha |^{2}}{4}(m_{3}^{2}+m_{2}^{2})+%
\frac{1}{2}\alpha _{R}km_{3}-\frac{1}{2}\alpha _{I}km_{2},
\end{equation}
\begin{equation}
b_{2}=\partial _{x}^{-1}\{-\frac{\gamma _{2}}{2ia_{2}^{\prime ^{2}}}-(A-\bar{%
A}+\bar{D}-D)\},
\end{equation}
where 
\[
\gamma _{1}=i\{\frac{1}{2}k^{2}\tau +\frac{|\alpha |^{2}}{2}%
(m_{3}km_{1}+m_{2}k_{y})- 
\]
\begin{equation}
\frac{1}{2}\alpha _{R}(k^{2}m_{1}+m_{3}k\tau +m_{2}k_{x})+\frac{1}{2}\alpha
_{I}[k(2k_{y}-m_{3x})-k_{x}m_{3}]\},
\end{equation}
\[
\gamma _{2}=-i\{\frac{1}{2}k^{2}\tau +\frac{|\alpha |^{2}}{2}%
(m_{3}km_{1}+m_{2}k_{y})+ 
\]
\begin{equation}
\frac{1}{2}\alpha _{R}(k^{2}m_{1}+m_{3}k\tau +m_{2}k_{x})+\frac{1}{2}\alpha
_{I}[k(2k_{y}-m_{3x})-k_{x}m_{3}]\}.
\end{equation}
Here $\alpha =\alpha _{R}+i\alpha _{I}$. In this case, $q,p$ satisfy the
(2+1)-dimensional equations of the nonlinear Schr\"{o}dinger (NLS) type [27] 
\begin{equation}
iq_{t}+(1+b)q_{\xi \xi }-bq_{\eta \eta }+vq=0,
\end{equation}
\begin{equation}
ip_{t}-(1+b)p_{\xi \xi }+bp_{\eta \eta }-vp=0,
\end{equation}
\begin{equation}
v_{\xi \eta }=-2\{(1+b)(pq)_{\xi \xi }-b(pq)_{\eta \eta }\}.
\end{equation}
These equations are the geometrical equivalent counterpart of the spin
system (54)-(55). Therefore, the spin system and the (2+1)-dimensional NLS
equations (105)-(107) turn out to be mutually geometrical equivalent.

\subsection{Gauge equivalence}

Now we prove that the spin system (54)-(55) and equations (105)-(107) are
not only equivalent in the geometrical sense, but are also each other gauge
equivalent. To this purpose, let us perform the gauge transformation $\Psi
=g\Phi $, where the function $\Phi $ is the solution of equations (59)-(60)
and $g$ is a 2x2 matrix such that 
\begin{equation}
S=g^{-1}\sigma _{3}g,
\end{equation}
and 
\begin{equation}
g_{Z^{+}}g^{-1}-\sigma _{3}g_{Z^{-}}g^{-1}=\left( 
\begin{array}{cc}
0 & q \\ 
p & 0
\end{array}
\right) .
\end{equation}
Under this transformation the function $\Psi $ obeys the following set of
linear equations 
\begin{equation}
\Psi _{Z^{+}}=\sigma _{3}\Psi _{Z^{-}}+B_{0}\Psi ,
\end{equation}
\begin{equation}
\Psi _{t}=4iC_{2}\Psi _{Z^{-}Z^{-}}+2C_{1}\Psi _{Z^{-}}+C_{0}\Psi ,
\end{equation}
where $B_{0},C_{j}$ are given by 
\begin{equation}
B_{0}=\pmatrix{ 0 & q \cr p & 0 },\quad C_{2}=\pmatrix{ b+1 & 0 \cr 0 & b }%
,\quad C_{1}=\pmatrix{ 0 & iq \cr ip & 0 },\quad C_{0}=\pmatrix{ c_{11} &
c_{12} \cr c_{21} & c_{22} },
\end{equation}

\bigskip and the functions $c_{ij}$ $(i,j=1,2)$ fulfills the equations

\begin{equation}
c_{12}=i[(4b+3)q_{Z^{-}}+q_{Z^{+}}],\quad
c_{21}=-i[(4b+1)p_{Z^{-}}+p_{Z^{+}}],
\end{equation}
\begin{equation}
c_{11Z^{-}}-c_{11Z^{+}}=i[(4b+3)(pq)_{Z^{-}}+(pq)_{Z^{+}}],
\end{equation}
\begin{equation}
c_{22Z^{-}}+c_{22Z^{+}}=i[(4b+1)(pq)_{Z^{-}}+(pq)_{Z^{+}}],
\end{equation}

\bigskip with $v=i(c_{22}-c_{11})$ (see (105)-(107)).

The compatibility condition of Eqs. (110)-(11) gives the equations
(105)-(107). This means that the spin model (54)-(55) and the NLS - type
equations (105)-(107) are gauge equivalent each other. Moreover, it is easy
to check that if $g$ satisfies equations Eq. (109), then $S$ given by (108)
satisfies Eqs. (54)-(55) with 
\begin{equation}
u=-2\ln \det g.
\end{equation}

\subsubsection{Reductions}

Equations (105)-(107), as its and equivalent spin system (54)-(55), contain
several integrable cases, namely

$i)$ $b=0$. Equations (105)-(107) yield the equations [27] 
\begin{equation}
iq_{t}+q_{\xi \xi }+vq=0,
\end{equation}
\begin{equation}
ip_{t}-p_{\xi \xi }-vp=0,
\end{equation}
\begin{equation}
v_{\eta }=-2(pq)_{\xi }.
\end{equation}

$\bigskip $

$ii)$ $a=b=-\frac{1}{2}$. Then, we give the Davey-Stewartson (DS) equation
[27] 
\begin{equation}
iq_{t}+q_{xx}+\alpha ^{2}q_{yy}+vq=0,
\end{equation}
\begin{equation}
ip_{t}-p_{xx}-\alpha ^{2}p_{yy}-vq=0,
\end{equation}
\begin{equation}
v_{xx}-\alpha ^{2}v_{yy}=2\{(pq)_{xx}+\alpha ^{2}(pq)_{yy}\},
\end{equation}
where $x=\xi -\eta ,y=\alpha (\xi +\eta )$.

$\bigskip $

$iii)$ Putting $b=0,\eta =t$, Eqs. (105)-(107) reduce to the
(1+1)-dimensional Ma [28] or Yajima-Oikawa [29] equations 
\begin{equation}
iq_{t}+q_{\xi \xi }+vq=0,
\end{equation}
\begin{equation}
ip_{t}-p_{\xi \xi }-vp=0,
\end{equation}
\begin{equation}
v_{t}+2(pq)_{\xi }=0,
\end{equation}
which are known to be integrable.

\section{Solutions of the spin system}

It could be of interest to study Eqs. (54)-(55) by the IST method. However,
to look for some special solutions, it is convenient to exploit the Hirota
bilinear method. To this aim, let us build up the bilinear form of (54)-(55)
for the compact case.

We obtain 
\begin{equation}
S^{+}=S_{1}+iS_{2}=\frac{2\bar{f}g}{\bar{f}f+\bar{g}g},\quad S_{3}=\frac{%
\bar{f}f-\bar{g}g}{\bar{f}f+\bar{g}g},
\end{equation}
\begin{equation}
u_{\xi }=-2i\frac{D_{\xi }(\bar{f}\circ f+\bar{g}\circ g)}{\bar{f}f+\bar{g}g}%
,\quad u_{\eta }=2i\frac{D_{\eta }(\bar{f}\circ f+\bar{g}\circ g)}{\bar{f}f+%
\bar{g}g},
\end{equation}

\bigskip where $S_{j}$ $(j=1,2,3)$ are the components of spin vector ${\bf S,%
}$ $S^{\pm }$ $=S_{1}\pm iS_{2},$and $u$ is the scalar potential.

Hence, from (116) we get 
\begin{equation}
u(\xi ,\eta ,t)=-2\ln (\mid f\mid ^{2}+\mid g\mid ^{2}).
\end{equation}

Substituting formulae (126) and (127) into the spin system (54)-(55), we
obtain the bilinear equations 
\begin{equation}
\lbrack iD_{t}-(b+1)D_{\xi }^{2}+bD_{\eta }^{2}](\bar{f}\circ g)=0,
\end{equation}
\begin{equation}
\lbrack iD_{t}-(b+1)D_{\xi }^{2}+bD_{\eta }^{2}](\bar{f}\circ f-\bar{g}\circ
g)=0,
\end{equation}
\begin{equation}
\{D_{\xi }D_{\eta }+D_{\eta }D_{\xi }\}(\bar{f}f+\bar{g}g)\circ (\bar{f}f+%
\bar{g}g)=0.
\end{equation}

\bigskip Equation (131) coincides with the compatibility condition $u_{\xi
\eta }=u_{\eta \xi }$.

Now we can construct some special solutions of Eqs. (54)-(55). In
particular, to construct vortex solutions, we start from Eqs. (129)-(130)
and assume that 
\begin{equation}
f=f(\xi ,t),\quad g=g(\xi ,t).
\end{equation}
Then Eq. (131) is satisfied automatically. At the same time, Eqs.
(129)-(130) are fulfilled if 
\begin{equation}
if_{t}+(b+1)f_{\xi \xi }=0\quad ig_{t}+(b+1)g_{\xi \xi }=0.
\end{equation}
Consequently, we are led to the following multi-vortex solutions 
\begin{equation}
g_{N}=\sum_{j=0}^{N}\sum_{m+2n=j}\frac{a_{j}}{m!n!}(\frac{2}{b+1})^{\frac{m}{%
2}}\xi ^{m}(2it)^{n},
\end{equation}
\begin{equation}
f_{N}=\sum_{j=0}^{N-1}\sum_{m+2n=j}\frac{b_{j}}{m!n!}(\frac{2}{b+1})^{\frac{m%
}{2}}\xi ^{m}(2it)^{n},
\end{equation}
where $a_{j}$ and $b_{j}$ are arbitrary complex constants, and $m,n$ are
non-negative integer numbers. In particular, the 1-vortex solution can be
derived by choosing 
\begin{equation}
f=b_{0},\quad g=a_{1}^{\prime }\xi +a_{0},
\end{equation}
where $a_{1}^{\prime }=a_{1}(\frac{2}{b+1})^{\frac{1}{2}}$.

So, the 1-vortex solution is static. To find a dynamic solution, we have to
consider the $N$-vortex solution using the forms 
\begin{equation}
f(\xi ,t)=b_{0}\prod_{j=1}^{N}[\xi -p_{j}(t)],
\end{equation}
\begin{equation}
g(\xi ,t)=a_{0}\prod_{j=1}^{N}[\xi -q_{j}(t)],
\end{equation}
where $p_{j}$ and $q_{j}$ denote the positions of the zeros of $f$ and $g$,
and $a_{0},b_{0}$ are constants. The evolution of $p_{j}$ and $q_{j}$ is 
\begin{equation}
p_{jt}=-i(b+1)\sum_{k\not=j}^{N}\frac{1}{p_{j}-p_{k}},
\end{equation}
\begin{equation}
q_{jt}=-i(b+1)\sum_{k\not=j}^{N}\frac{1}{q_{j}-q_{k}},
\end{equation}
where $j,k=1,2,...,N$. These equations are related to the Calogero-Moser
system.

\section{Special surfaces corresponding to vortex solutions of the spin
system}

\bigskip

This Section is devoted to the construction of explicit surfaces. To this
aim, let us start from the 1-vortex solution of the spin system (54)-(55).
By choosing for simplicity $E=1$, Eqs. (41)-(42) become 
\begin{equation}
k=L,\quad \sigma =\frac{F_{x}}{\sqrt{g}},\quad \tau =\frac{M-LF}{\sqrt{g}},
\end{equation}
\begin{equation}
m_{1}=\frac{N-MF}{\sqrt{g}},\quad m_{2}=-\frac{G_{x}}{2\sqrt{g}},\quad
m_{3}=M.
\end{equation}

On the other hand, from (38) and (89) we get 
\begin{equation}
{\bf r}_{x}(\xi ,\eta ,t)={\bf S}(\xi ,\eta ,t).
\end{equation}

\bigskip

Now let us consider by way of example the surface associated with the
1-vortex solution of the Ishimori system, whose components are

\begin{eqnarray*}
&& 
\begin{array}{ll}
S_{3}=\frac{1-b^{2}\left| \Xi \right| ^{2}}{1+b^{2}\left| \Xi \right| ^{2}},
& S^{+}=2be^{i\delta }\frac{\Xi }{1+b^{2}\left| \Xi \right| ^{2}},
\end{array}
\\
&&\quad u=2\left( \ln \left( b\right) +\ln \left[ 1+b^{2}\left| \Xi \right|
^{2}\right] \right) ,
\end{eqnarray*}
where $\Xi $ denotes the complex variable $\Xi =2a\exp \left( i\gamma
\right) \left( x+iy\right) +1$, $a,b,$ $\gamma $ and $\delta $ being real
constants.

Then, by resorting to the formula $e_{1}=r_{x}/\sqrt{E}$(with $E\equiv 1)$,
we can integrate to yield the following components for the position vector 
\begin{equation}
\begin{array}{c}
r_{1}={\frac{\hat{c}\,\ln (1+{b^{2}}\left| \Xi \right| ^{2})}{a\,b}}-{\frac{%
\sqrt{2}\,\arctan \left[ \,b\,\sqrt{{\frac{2}{\Omega }}}\,\left(
a\,x+c\right) \right] ~\left( \tilde{c}-2\,a\,\hat{s}y\right) }{a\,\sqrt{%
\Omega }},} \\ 
r_{2}={\frac{\hat{s}\ln (1+{b^{2}}\,\left| \Xi \right| ^{2})\,}{a\,b}}-{%
\frac{\sqrt{2}\,\arctan \left[ \,b\,\sqrt{{\frac{2}{\Omega }}}\,\left(
a\,x+c\right) \right] ~\left( 2\,a\,\hat{c}y\,+\tilde{s}\right) }{a\,\sqrt{%
\Omega }},} \\ 
r_{3}=-x+{\frac{2\,\sqrt{2}\,\arctan \left[ \,b\,\sqrt{{\frac{2}{\Omega }}}%
\,\left( a\,x+c\right) \right] ~}{a\,b\,\sqrt{\Omega }},}
\end{array}
\label{position}
\end{equation}
where for the sake of clarity we have introduced the second degree
polynomial $\Omega =2+b^{2}\left[ 1+2a^{2}y^{2}-\cos \left( 2\gamma \right)
+4a\sin \left( \gamma \right) y\right] $ and the constants $c=\cos \left(
\gamma \right) ,$ $s=\sin \left( \gamma \right) ,$ $\hat{c}=$ $\cos \left(
\gamma +\delta \right) ,$ $\hat{s}=$ $\sin \left( \gamma +\delta \right) ,$ $%
\tilde{c}=-\cos (\delta )+\cos (2\,\gamma +\delta ),$and $\tilde{s}=-\sin
(\delta )+\sin (2\,\gamma +\delta ).$ Furthermore, we have put identically
equal to zero any arbitrary function of integration in $y$ only.

From them, with the help of the various formulae given in Section 3, we
obtain the coefficients of the I-fundamental form 
\begin{eqnarray*}
&& 
\begin{array}{ll}
E=1, & F={\frac{4\,b\,\left( s+a\,y\right) \,\left( b\,\left( c+a\,x\right)
\,\sqrt{\Omega }+\sqrt{2}\,\left( 1+{b^{2}}\,\left| {\Xi }\right| {^{2}}%
\right) \,At\left[ x,y\right] \right) }{\left( 1+{b^{2}}\,\left| {\Xi }%
\right| {^{2}}\right) \,{{\Omega }^{{\frac{3}{2}}}}}}
\end{array}
\\
&& 
\begin{array}{c}
G={\frac{2\,\left( 4\,{b^{2}}\,{{\left( s+a\,y\right) }^{2}}\,\Omega
+8\,\left( 1+{b^{2}}\,\left| {\Xi }\right| {^{2}}\right) \,At\left[ x,y%
\right] {^{2}}\right) }{\left( 1+{b^{2}}\,\left| {\Xi }\right| {^{2}}\right)
\,{{\Omega }^{2}}}}
\end{array}
\end{eqnarray*}
where $At[x,y]=\arctan \left[ \,b\,\sqrt{{\frac{2}{\Omega }}}\,\left(
a\,x+c\right) \right] ,$ and analogously for the II-fundamental form

\[
\begin{array}{l}
\sqrt{g}L=\frac{8\,a\,b\,\left[ -\left( {b^{3}}\,\left( c+a\,x\right) \,{{%
\left( s+a\,y\right) }^{2}}\,\sqrt{\Omega }\right) +\sqrt{2}\,\left( 1+{b^{2}%
}\,\left| {\Xi }\right| {^{2}}\right) \,At\left[ x,y\right] \right] }{{{%
\left( 1+{b^{2}}\,\left| {\Xi }\right| {^{2}}\right) }^{2}}\,{{\Omega }^{{%
\frac{3}{2}}}}},\quad \sqrt{g}M={\frac{-4\,a\,{b^{2}}\,\left( s+a\,y\right) 
}{{{\left( 1+{b^{2}}\,\left| {\Xi }\right| {^{2}}\right) }^{2}}}} \\ 
\sqrt{g}N={\frac{8\,a\,b\,\left( 2\,\sqrt{2}\,\left( 1+{b^{2}}\,\left| {\Xi }%
\right| {^{2}}\right) \,\sqrt{\Omega }\,\,At\left[ x,y\right] +8\,b\,\left(
c+a\,x\right) \,\left( 1+{b^{2}}\,\left| {\Xi }\right| {^{2}}\right) \,\,At%
\left[ x,y\right] {^{2}}+{b^{3}}\,\left( c+a\,x\right) \,{{\left(
s+a\,y\right) }^{2}}\,\Omega \,\left( 2+\Omega \right) \right) }{{{\left( 1+{%
b^{2}}\left| \,{\Xi }\right| {^{2}}\right) }^{2}}\,{{\Omega }^{3}}}}
\end{array}
\]
where $g$ in the metric factor $\sqrt{g}$ is expressed by 
\[
\begin{array}{c}
g=\frac{8}{{{\left( 1+{b^{2}}\,\left| {\Xi }\right| {^{2}}\right) }^{2}}\,{{%
\Omega }^{3}}}\{\,{b^{2}}\,{{\left( s+a\,y\right) }^{2}}\,\Omega \,-2\,{b^{2}%
}\,{{\left( c+a\,x\right) }^{2}}+\left( 1+{b^{2}}\,\left| {\Xi }\right| {^{2}%
}\right) \,\Omega ) \\ 
-4\,\sqrt{2}\,{b^{3}}\,\left( c+a\,x\right) \,{{\left( s+a\,y\right) }^{2}}%
\,\left( 1+{b^{2}}\,\left| {\Xi }\right| {^{2}}\right) \,\sqrt{\Omega }\,At%
\left[ x,y\right] +4\,{{\left( 1+{b^{2}}\,\left| {\Xi }\right| {^{2}}\right) 
}^{2}}\,At\left[ x,y\right] {^{2}\}.} \\ 
.
\end{array}
\]
The Gauss curvature $K$ and the mean curvature $H$ are given by (see (32)
and (33)) are given by 
\[
K=\frac{8\,{a^{2}}\,{b^{2}}\,}{{{\left( 1+{b^{2}}\,{{\Xi }^{2}}\right) }^{4}}%
\,{{\Omega }^{{\frac{9}{2}}}}}{\{}-{b^{2}}\,{{\left( s+a\,y\right) }^{2}}\,{{%
\Omega }^{{\frac{3}{2}}}}\,\left( {{\Omega }^{3}}+4\,{b^{4}}\,{{\left(
c+a\,x\right) }^{2}}\,{{\left( s+a\,y\right) }^{2}}\,\left( 2+\Omega \right)
\right) 
\]

\[
+4\sqrt{2}b^{3}(c+ax)(s+ay)^{2}(1+b^{2}\mid \Xi \mid ^{2})\Omega
^{2}At[x,y]- 
\]

\[
16(1+b^{2}\mid \Xi \mid ^{2})[2\,{b^{4}}\,{{\left( c+a\,x\right) }^{2}}\,{{%
\left( s+a\,y\right) }^{2}}-(1+b^{2}\mid \Xi \mid ^{2})]\sqrt{\Omega }\,{{%
At[x,y]}^{2}}+ 
\]

\[
32\sqrt{2}b(c+ax)(1+b^{2}\mid \Xi \mid ^{2})^{2}{{At[x,y]}^{3}\},} 
\]

\bigskip 
\[
H=\frac{4ab}{\left( 1+{b^{2}}\,{{\left| {\Xi }\right| }^{2}}\right)
^{3}\Omega ^{\frac{7}{2}}\sqrt{g}}\ \{{b^{3}}\,\left( c+a\,x\right) \,{{%
\left( s+a\,y\right) }^{2}}\,{{\Omega }^{{\frac{3}{2}}}}\,[\left( 1+{b^{2}}\,%
{{\left| {\Xi }\right| }^{2}}\right) \,\left( 2+\Omega \right) + 
\]

\[
4\,(2\,{b^{2}}\,{{\left( s+a\,y\right) }^{2}}+\Omega )] 
\]

\[
+2\,\sqrt{2}\,\left( 1+{b^{2}}\,{{\left| {\Xi }\right| }^{2}}\right)
\,\Omega \,\left( 1-4\,{b^{2}}\,{{\left( s+a\,y\right) }^{2}}+{b^{2}}\,{{%
\left| {\Xi }\right| }^{2}}+2\,{b^{2}}\,{{\left( s+a\,y\right) }^{2}}%
\,\Omega \right) \,At[x,y]+ 
\]

\[
8\,b\,\left( c+a\,x\right) \,\left( 1+{b^{2}}\,{{\left| {\Xi }\right| }^{2}}%
\right) \,\left( 1+2\,{b^{2}}\,{{\left( s+a\,y\right) }^{2}}+{b^{2}}\,{{%
\left| {\Xi }\right| }^{2}}\right) \sqrt{\Omega }\,At[{{x,y]}^{2}}- 
\]

\[
16\,\sqrt{2}\,{{\left( 1+{b^{2}}\,{{\left| {\Xi }\right| }^{2}}\right) }^{2}}%
\,At[{{x,y]}^{3}\},} 
\]

respectively.

An example of a surface associated with the 1-vortex solution of the
Ishimori system is drawn in Fig 1.

\section{Conclusions}

\bigskip

In this paper we have established some notable connections among the purely
topological CS theory, deformations of surfaces, and integrable equations in
(2+1)-dimensions. However, many questions remain open and deserve further
investigation, such as for example the search for other integrable classes
of deformations of surfaces, the determination of the Hamiltonian structure
and the possible interpretation of the solutions by a physical point of
view. To this regard, in particular we have found exact vortex solutions of
the (2+1)-dimensional spin system. {\bf \ } Furthermore, we have seen that
the dynamics of vortices is governed by a system of the Calogero-Moser type.
To conclude, we notice that another approach exists to study integrable
(2+1)-dimensional deformations of surfaces, i.e. the method developed mainly
by Konopelchenko, Taimanov and coworkers [3,30]. The essential tool of their
procedure is the use of a generalized Weierstrass representation for a
conformal immersion of surfaces into $R^{3}$ or $R^{4},$ together with a
linear problem related to this representation. The method devised in [3,30]
allows one to express integrable deformations of surfaces via hierarchies of
integrable equations, such as the Nizhnik-Veselov-Novikov and the DS
equations, and so on. We think that our approach and that described in [3]
should be pursued in parallel, with the purpose to achieve possible
complementary results on the link between integrable deformations of
surfaces and completely integrable partial differential equations.

\section{Acknowledgments}

The authors are grateful to V.S. Dryuma and B.G. Konopelchenko for very
helpful discussions. This work was supported in part by MURST of Italy,
INFN-Sezione di Lecce and INTAS (grant 99-1782). One of the authors (R.M.)
thanks the Department of Physics of the Lecce University for its warm
hospitality.

%-------------- F I G U R E    1
\leavevmode
\begin{figure}[ht]
%\vspace{-5in}
\hspace{5in}
\epsfxsize=5.2in
\epsfbox{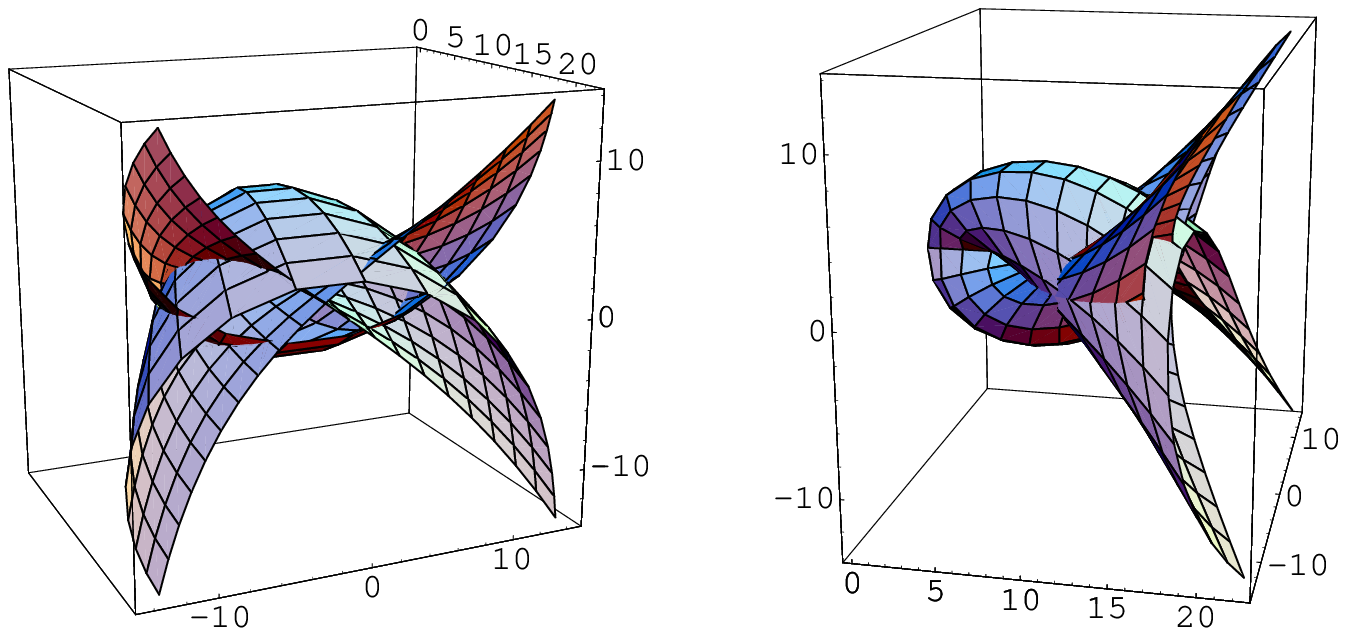}
\vspace{-0.0in}
\caption{{We drawed the same surface from two different points of view. The parameters used in eq. (144) are $ a=1, b = 0.1, \gamma = \delta = 0 $ }}
\end{figure}
\leavevmode

\bigskip

\begin{thebibliography}{99}
\bibitem{1}  A. Sym, O. Ragnisco, D. Levi and M. Bruschi, Lett. Nuovo
Cimento {\bf 44}, 529 (1991); A.I. Bobenko, ''Surfaces in terms of 2 by 2
matrices, old and new integrable cases'', in {\it Harmonic Maps and
Integrable Systems}, edited by A.P. Fordy and J.C. Wood, Aspects of
Mathematics (Friedr. Vieweg and Sohn, 1994), p. 83.

\bibitem{2}  O. Ceyhan, A.S. Fokas and M. Gurses, J. Math. Phys. {\bf 41},
2251 (2000).

\bibitem{3}  B.G. Konopelchenko, {\it Stud. Appl. Math. {\bf 96},} 9 (1996).

\bibitem{4}  G. Darboux, {\it Le\c{c}on sur la th\'{e}orie g\'{e}n\'{e}rale
des surfaces}, vol. 4 (Paris, Gauthier-Villars, 1910).

\bibitem{5}  G. Darboux, {\it Le\c{c}on sur les syst\`{e}me orthogonaux et
les coorden\'{e}es curvilignes }(Paris, Gauthier-Villars, 1910).

\bibitem{6}  V.E. Zakharov, Duke Math. Journ. {\bf 94}, 103 (1998); V.E.
Zakharov and S.E. Manakov, Dok. Math. {\bf 57}, 471 (1998).

\bibitem{7}  R. Dijkgraaf, E. Verlinde and H. Verlinde, Nucl. Phys. B {\bf %
352}, 59 (1991); E. Witten, Nucl. Phys. B {\bf 340}, 289 (1990).

\bibitem{8}  L. Martina, O.K. Pashaev and G. Soliani, J. Math. Phys. {\bf 38}%
, 1397 (1997).

\bibitem{9}  S. Deser, R. Jackiw and S. Templeton, Ann. Phys. {\bf 180}, 372
(1982).

\bibitem{10}  B.A. Dubrovin, A.T. Fomenko and S.P. Novikov, {\it Modern
Differential Geometry} (Springer, Berlin, 1984).

\bibitem{11}  E. Witten, Commun. Math. Phys. {\bf 121}, 351 (1989).

\bibitem{12}  {\it The Quantum Hall Effect}, edited by S. Girvin and R.
Prange (Springer, New York, 1990).

\bibitem{13}  L. Martina, O.K. Pashaev and G. Soliani, Phys. Rev. D{\bf \ 58}%
, 084025 (1998).

\bibitem{14}  G. Dunne, {\it Self-dual Chern-Simons theories }(Springer,
Berlin, 1995).

\bibitem{15}  L. Martina, O.K. Pashaev and G. Soliani, Mod. Phys. Lett. A%
{\bf \ 8}, 34 (1993).

\bibitem{16}  V.E. Zakharov and A.B. Shabat, Funct. Anal. Appl. {\bf 13},
166 (1979); B.G. Konopelchenko, {\it Introduction to Muldimensional
Integrable Equations} (Plenum, New York, 1992).

\bibitem{17}  V.D. Lipovski and A.V. Shirokov, Funct. Anal. and Appl. {\bf 23%
}, 225 (1990).

\bibitem{18}  R.A. Leo, L. Martina and G. Soliani, J. Math. Phys.{\bf \ 33},
1515 (1992).

\bibitem{19}  T. Eguchi, P.B. Gilkey and A.J. Hanson, Phys. Rep. {\bf 66},
213 (1980).

\bibitem{20}  Y. Ishimori, {Prog. Theor. Phys.} {\bf 72,} 33 (1984).

\bibitem{21}  R. Myrzakulov, S. Vijayalakshmi, G.N. Nugmanova and M.
Lakshmanan, Phys. Lett. A {\bf 233, }391 (1997).

\bibitem{22}  R. Myrzakulov, S. Vijayalakshmi, R.N. Syzdykova and M.
Lakshmanan, J. Math. Phys. {\bf 39,} 2122 (1998).

\bibitem{23}  M. Lakshmanan, R. Myrzakulov, S. Vijayalakshmi and A.K.
Danlybaeva , J. Math. Phys. {\bf 39.} 3765 (1998).

\bibitem{24}  R. Myrzakulov, G.N. Nugmanova and R.N. Syzdykova, J. Phys. A:
Math. Gen. {\bf 31,} 9535 (1998).

\bibitem{25}  R. Myrzakulov, A.K. Danlybaeva and G.N. Nugmanova, Theor.
Math. Phys. {\bf 118}, 347 (1999).

\bibitem{26}  R. Myrzakulov, M. Daniel and R. Amuda, Physica A {\bf 234},
715 (1997).

\bibitem{27}  V.E. Zakharov, in: {\it Solitons, } editors R. K. Bullough and
P.J. Caudrey (Springer, Berlin 1980).

\bibitem{28}  Y.C. Ma, Stud. Appl. Math. {\bf 59,} 201 (1978).

\bibitem{29}  N. Yajima and M. Oikawa, Prog. Theor. Phys. {\bf 56}, 1719
(1976).

\bibitem{30}  I.A. Taimanov, Siberian Math. Journal {\bf 40}, 1146 (1999).
\end{thebibliography}
\end{document}